\documentclass[8.5pt,twoside]{article}

\usepackage[super,sort&compress,comma]{natbib} 
\usepackage{times,mathptm}
\usepackage{sectsty}
\usepackage{balance} 
\usepackage{natbib}

\usepackage[text={11.3cm,20.4cm},centering]{geometry} 

\usepackage{graphicx} %eps figures can be used instead
\usepackage{lastpage}
\usepackage[format=plain,justification=raggedright,singlelinecheck=false,font=small,labelfont=bf,labelsep=space]{caption} 
\usepackage{fancyhdr}
\begin{document}

\thispagestyle{plain}
\fancypagestyle{plain}{
%\fancyhead[L]{\includegraphics[height=8pt]{headers/LH}}
%\fancyhead[C]{\hspace{-1cm}\includegraphics[height=15pt]{headers/CH}}
%\fancyhead[R]{\includegraphics[height=10pt]{headers/RH}\vspace{-0.2cm}}
\renewcommand{\headrulewidth}{1pt}}
\renewcommand{\thefootnote}{\fnsymbol{footnote}}
\renewcommand\footnoterule{\vspace*{1pt}% 
\hrule width 11.3cm height 0.4pt \vspace*{5pt}} 
\setcounter{secnumdepth}{5}

\makeatletter 
\renewcommand{\fnum@figure}{\textbf{Fig.~\thefigure~~}}
\def\subsubsection{\@startsection{subsubsection}{3}{10pt}{-1.25ex plus -1ex minus -.1ex}{0ex plus 0ex}{\normalsize\bf}} 
\def\paragraph{\@startsection{paragraph}{4}{10pt}{-1.25ex plus -1ex minus -.1ex}{0ex plus 0ex}{\normalsize\textit}} 
\renewcommand\@biblabel[1]{#1}            
\renewcommand\@makefntext[1]% 
{\noindent\makebox[0pt][r]{\@thefnmark\,}#1}
\makeatother 
\sectionfont{\large}
\subsectionfont{\normalsize} 

\fancyfoot{}
%\fancyfoot[LO,RE]{\vspace{-7pt}\includegraphics[height=8pt]{headers/LF}}
%\fancyfoot[CO]{\vspace{-7pt}\hspace{5.9cm}\includegraphics[height=7pt]{headers/RF}}
%\fancyfoot[CE]{\vspace{-6.6pt}\hspace{-7.2cm}\includegraphics[height=7pt]{headers/RF}}
\fancyfoot[RO]{\scriptsize{\sffamily{1--\pageref{LastPage} ~\textbar  \hspace{2pt}\thepage}}}
\fancyfoot[LE]{\scriptsize{\sffamily{\thepage~\textbar\hspace{3.3cm} 1--\pageref{LastPage}}}}
\fancyhead{}
\renewcommand{\headrulewidth}{1pt} 
\renewcommand{\footrulewidth}{1pt}
\setlength{\arrayrulewidth}{1pt}
\setlength{\columnsep}{6.5mm}
\setlength\bibsep{1pt}

\noindent\LARGE{\textbf{Characterizing the chemical pathways for water formation --\\ A deep search for hydrogen peroxide\dag}}
\vspace{0.6cm}

\noindent\large{\textbf{B\'ereng\`ere Parise,$^{\ast}$\textit{$^{ab}$} Per Bergman,\textit{$^{c}$} and
Karl Menten\textit{$^{b}$}}}\vspace{0.5cm}

\noindent\textit{\small{\textbf{Received 29th November 2013, Accepted 11th February 2014\newline
First published on the web 2014}}}

\noindent \textbf{\small{DOI: 10.1039/c3fd00115f}}
\vspace{0.6cm}

\footnotetext{\textit{$^{a}$~School of Physics and Astronomy, University of Cardiff, The Parade, CF24 3AA, Cardiff, UK. Fax: +44 (0)29 208 74056; Tel: +44 (0)29 208 74649; E-mail: Berengere.Parise@astro.cf.ac.uk}}
\footnotetext{\textit{$^{b}$~Max-Planck-Institut f\"ur Radioastronomie, Auf dem H\"ugel 69, 53121 Bonn, Germany. }}
\footnotetext{\textit{$^{c}$~Onsala Space Observatory, Chalmers University of Technology, 439 92 Onsala, Sweden. }}

\footnotetext{\dag~Based on observations with the Atacama Pathfinder EXperiment (APEX) telescope. APEX is a collaboration between the Max Planck Institute for Radio Astronomy, the European Southern Observatory, and the Onsala Space Observatory.}

\noindent \normalsize{
In 2011, hydrogen peroxide (HOOH) was observed for the first time outside the solar system (Bergman \textit{et al., A\&A}, 2011, \textbf{531}, L8). This detection appeared \textit{a posteriori} quite natural, as HOOH is an intermediate product in the formation of water on the surface of dust grains. 
Following up on this detection, we present a search for HOOH in a diverse sample of sources in different environments, including low-mass protostars and regions with very high column densities, such as Infrared Dark Clouds (IRDCs). We do not detect the molecule in any other source than Oph A, and derive  
3$\sigma$ upper limits for the abundance of HOOH relative to H$_2$ lower than in Oph A for most sources. This result sheds a different light on our understanding of the detection of HOOH in Oph A, and
shifts the puzzle to why this source seems to be special. Therefore we rediscuss the detection of HOOH in Oph A, as well as the implications of the low abundance of HOOH, and its similarity with the case of O$_2$. 
Our chemical models show that the production of HOOH is extremely sensitive to the temperature, and favored only in the range 20$-$30\,K.
The relatively high abundance of HOOH observed in Oph A suggests that the bulk of the material lies at a temperature in the range 20$-$30\,K.
}
\vspace{0.5cm}

\section{Introduction}

Water has long been known to exist in star-forming regions, both in the gas phase \cite{Bergin00} and in the form of ices \cite{Whittet96}. The {\it Herschel Space Observatory} has recently shown that water is present virtually everywhere where it was looked for, and provided detections in new environments such as cold prestellar cores \cite{Caselli10, Caselli12} and protoplanetary disks \cite{Hogerheijde11} (see, e.g., the results of the WISH Key Project \cite{vanDishoeck11}). 
 
The relatively low abundance of water in the gas phase of cold regions, as well as its high abundance in 
the ices coating dust grains, has led to the hypothesis that water forms very efficiently on the surface of 
dust grains. Recent laboratory experiments have investigated the details of the 
different formation pathways \cite{Hiraoka98, Miyauchi08, Mokrane09, Cuppen10}. In these experiments, water was shown to form through three different
routes, starting with the hydrogenation of, respectively, O, O$_2$ and O$_3$. The relative importance of the three routes in the different environments is not yet fully understood, but can have huge impacts on the 
resulting composition of the formed ices.

HOOH is formed on the grains as a precursor of water in the O$_2$ route: 
\[ {\rm O_2 + H} \rightarrow {\rm HO_2}  ~~~~~~~~~~~~~~~~~~~~~~~(1)\]
\[ {\rm HO_2 + H} \rightarrow {\rm HOOH}  ~~~~~~~~~~~~~~~~(2)\]
\[ {\rm HOOH + H} \rightarrow {\rm H_2O + OH}    ~~~~~~(3)\]

Reactions (1) and (2) on dust grains were hypothesized by Allen et al.\cite{Allen77} based on the related
gas-phase reactions, and the addition of reaction (3) in models was proposed by Tielens et al.\cite{Tielens82}. Laboratory experiments recently showed that these reactions indeed proceed \cite{Miyauchi08, Cuppen10}.

In view of the ubiquity of water, and the central role of HOOH in its formation, it came somewhat 
as a surprise that HOOH was only recently detected for the first time in the interstellar medium\cite{Bergman11b}, 
namely in $\rho$\,Oph A.  
Subsequent detailed chemical modelling, solving gas-phase and grain surface chemistry with the 
Hybrid Moment Equation (HME) method \cite{Du11}, showed that HOOH is indeed a major precursor of H$_2$O in that source \cite{Du12}. 
Further confidence was brought to the modelling by the (predicted) detection of HO$_2$ \cite{Parise12a}, raising again the question of 
why HOOH had not been detected earlier.
No dedicated observational search had been reported previously though, although the frequencies were long known \cite{Helminger81,Petkie95}, so that the lack of detections may just be attributed to the lack of search efforts. 

Following the detection of HOOH towards Oph A, we intend here to clarify the observational picture, and present a search for HOOH in a sample of ten sources of different nature and in different environments.

\section{Observations}

\subsection{Technical details}
Using the APEX telescope \cite{Guesten06}, we targetted the HOOH 3$_{0,3}$--2$_{1,1}$ transition towards a sample of sources. The energy of the upper level for this transition is 31\,K. This line was the brightest of a set of transitions observed towards $\rho$ Oph A \cite{Bergman11b}. The sources are listed in Table\,\ref{tbl:sourcelist} and described in detail in Section \ref{sources}.
The observations were made on 2011 August 6 and 7, under acceptable (PWV\,=\,1$-$3\,mm) weather conditions. 
The APEX1 receiver \cite{Vassilev08} was tuned at the frequency of the HOOH line (219.166860 GHz) and was connected to the XFFTS spectrometer \cite{Klein12}. 
The main beam efficiency at this frequency is 0.75, and the angular resolution is 28$''$ (FWHM).

\subsection{Source sample} 
\label{sources}

In the following we give a brief description of our source sample (Table \ref{tbl:sourcelist}). The sources have 
been selected based on different criteria. An important condition was high column densities, to increase the 
sensitivity of the detection. Some sources were selected because they share some characteristics with Oph A: either proximity (Oph B) to test the role of the local conditions, or average temperature conditions around 20-25\,K (IRDC sources, envelopes of low-mass protostars). Finally, to span as many different chemical environments as possible, we also selected one prototypical source representing each of the following chemical classes: hot cores (NGC6334), hot corinos (NGC1333-IRAS4A), warm carbon chain chemistry (L1527), and ``hot chemistry without hot cores'' \cite{Requena06} (G1.6).

Relevant information (dust and gas temperatures, as well as H$_2$ column densities) for our sample is listed in Table \ref{tbl:sourceproperties}, and references for these values are given in the following text. We have
converted H$_2$ column densities found in the literature to correspond to the value averaged
in a 28$''$ beam, similar to our HOOH observations. The conversion is discussed source by source.
For comparison, we also include Oph A, where HOOH was detected \cite{Bergman11b}.

\begin{table}[!ht]
\small
\begin{centering}
  \caption{~Source list}
  \label{tbl:sourcelist}
  \begin{tabular}{l l l l } 
      \hline
    \noalign{\smallskip}
    Source & RA (J2000) & DEC  (J2000)&   Distance  $^{\rm (References)}$ \\
    & & & pc  \\
    \hline
    \noalign{\smallskip}
    $\rho$ Oph-A SM1& 16:26:27.2& $-$24:24:04 & 120 $^($\cite{Loinard08}$^)$\\
    \hline
    \noalign{\smallskip}
    $\rho$ Oph-B2-MM8       &   16:27:28.0   &  $-$24:27:06.9    &  120 $^($\cite{Loinard08}$^)$  \\

    G15.01-0.67         &   18:20:21.22 & $-$16:12:42.2    &    2100 $^($\cite{Chini08}$^)$\\    
     G018.82-00.28MM1    &   18:25:56.1   & $-$12:42:48      &  4800 $^($\cite{Simon06}$^)$  \\
    G018.82-00.28MM4    &   18:26:15.5   & $-$12:41:32      &  4800 $^($\cite{Simon06}$^)$  \\
    G028.53-00.25MM1A   &   18:44:18.08  &$-$03:59:34.33  &  5700 $^($\cite{Simon06}$^)$  \\ 
NGC6334I(N)         &   17:20:54.63 & $-$35:45:08.9    &   1600 $^($\cite{Persi08}$^)$\\
G1.6-0.025          &   17:49:43.6  & $-$27:33:52      &   8000 $^($\cite{Menten09,Reid93}$^)$\\ 
NGC1333-IRAS4A       &   03:29:10.3  & $+$31:13:32      &   235 $^($\cite{Hirota08}$^)$ \\
L1527               &   04:39:53.9   &$+$26:03:10      &  140 $^($\cite{Kenyon94}$^)$ \\
RCrA-IRS7B           &   19:01:56.4  & $-$36:57:27      &  130  $^($\cite{Neuhauser08}$^)$  \\
    \hline
  \end{tabular}\\
  \end{centering}
 \end{table}

{\bf $\rho$ Oph-B2-MM8} is located in the same dark cloud system as $\rho$ Oph A. It is the brightest 
clump in the Oph B2 region at 1.3\,mm, as observed by \citet{Motte98}.
These authors estimated a mass of 1.5\,M$_\odot$ assuming a dust temperature of 12\,K\cite{Ristorcelli97}.
The peak column density in a 11$''$ beam for the Oph B2 region is 4.1$\times$10$^{23}$ cm$^{-2}$ 
(\citet{Motte98}), and that should correspond to the value centered on MM8. 
They quote a source size of 4000 $\times$ 4000 AU, which corresponds to 25$''$ at the distance they adopted at that time (160\,pc). We can therefore extrapolate the H$_2$ column density in a 28$''$ beam to be 2.2\,$\times$\,10$^{23}$\,cm$^{-2}$.

{\bf G15.01-0.67} is located in the M17 nebula (also known as the Omega Nebula). It is a SCAMPS source (high-mass pre/protocluster clump detected in the SCUBA Massive Pre/Protocluster core Survey, \citet{Thompson06}) studied by \citet{Pillai07}. It was detected on the edge of one of the SCUBA fields from \citet{Thompson06}. C$^{18}$O excitation temperature is 32\,K, and the NH$_3$ rotational temperature 26\,K \cite{Pillai07}.
The total $N$(H$_2$) column density is 16.6$\times$10$^{23}$ cm$^{-2}$, as derived from the 850\,$\mu$m dust continuum flux smoothed to a resolution of 20$''$\cite{Pillai07}. In the extreme case of a point source,
the column density should be reduced by a factor of 2 to obtain it within a 28$''$ beam. If the source has
a 20$''$ size, the decrease would be by a factor 1.5, and this is what we adopt here.

{\bf G018.82-00.28 MM1 and MM4} are the two most massive cores within the IRDC MSXDC G018.82-00.28.
They were detected and characterised by \citet{Rathborne06} using the MAMBO 1.2\,mm continuum receiver. Their angular sizes are 23$''$ and 31$''$ respectively \cite{Rathborne06}.
Their dust temperatures were derived from broadband SEDs (resp. 26 and 17\,K) by 
\citet{Rathborne10}, and their masses were then determined from the 1.2\,mm flux (resp. 495 and 228 M$_{\odot}$). The H$_2$ column densities in the MAMBO 11$''$ beam are 3.2 and 1.5 $\times$10$^{23}$ cm$^{-2}$, respectively. Taking into account their angular sizes, this leads to column densities
of 1.6 and 0.93 $\times$10$^{23}$ cm$^{-2}$, respectively, when averaged in a 28$''$ beam.
We find that the LSR velocity from the MM1 core as measured from the H$_2$CO lines (40.4 km/s) is surprisingly different
from that of the rest of the complex, as measured by \citet{Rathborne06} (65.8 km/s). This could be a sign 
that the MM1 core, which already appears spatially separated from the rest of the complex on MSX images \cite{Rathborne06}, is actually at a different distance than the rest of the complex.

\begin{table}[!ht]
\small
\begin{centering}
  \caption{~Source properties}
  \label{tbl:sourceproperties}
  \begin{tabular}{l l c cc ll} 
      \hline 
    \noalign{\smallskip}

    Source   &    $v_{\rm lsr}$    &    Linewidth    &  $T_{\rm dust}$  &  $T_{\rm gas}$ & $T_{\rm H_2CO}$$^c$  & $N$(H$_2$)$^e$   \\
    & km/s & km/s &  K & K & K & cm$^{-2}$ \\
    \hline 
    \noalign{\smallskip}
 $\rho$ Oph-A SM1 &  &  &  & 24 &  33$\pm$3 &  1.5$\times$10$^{23}$$^{(f)}$ \\   
    \hline 
    \noalign{\smallskip}
  $\rho$ Oph-B2-MM8         &     3.9$^a$    &  1.1$\pm$0.1$^a$  &   12  & & $\le$\,16  & 2.2$\times$10$^{23}$ \\
    G15.01-0.67         &   18.4$^b$  &  4.9$\pm$0.1$^a$  &   & 26\,--\,32 & 64$\pm$11 & 1.1$\times$10$^{24}$    \\ 
    G018.82-00.28MM1    &      40.4$^a$  &  5.7$\pm$0.1$^a$  &  26  &  &  61$\pm$10 & 1.6$\times$10$^{23}$\\
    G018.82-00.28MM4    &     64.9$^a$  &  4.4$\pm$0.3$^a$ & 17 & & $\le$\,29 & 9.3$\times$10$^{22}$ \\
G028.53-00.25MM1   &       86.3$^a$    & 6.34$\pm$0.2$^a$ & 17 & & 57$\pm$9 & 3.6$\times$10$^{23}$\\
NGC6334I(N)         &      -3.8$^b$   &    5.7$\pm$0.1$^b$ &  30\,--\,35 &  & 160$\pm$67 & 8.5$\times$10$^{23}$  \\
G1.6-0.025          &     51.7$^b$  & 5.3$\pm$0.3$^b$ & & 60 & 195$\pm$107& 4$\times$10$^{22}$\\
NGC1333-IRAS4A       &      7.0$^a$ & $\le$3.3$^{d}$ & &  24 & --$^d$  &1.3$\times$10$^{23}$\\
L1527               &       5.9$^a$    & 1.1$\pm$0.1$^a$ & & 16 & 23$\pm$2 &  4.1$\times$10$^{22}$\\
RCrA-IRS7B           &      5.7$^a$ & 2.5$\pm$0.2$^a$  & & 22\,$-$\,40& 47$\pm$6 & 5.9$\times$10$^{22}$ \\    \hline
  \end{tabular}\\
  \end{centering}
  $^a$ measured on the H$_2$CO 218.222\,GHz low-energy line\\
  $^b$ measured on the H$_2$CO 218.475\,GHz line, because the low-energy line is double peaked (NGC6334I(N)) or shows other signs of high opacity (G1.6-0.025). \\
  $^c$ H$_2$CO rotational temperature deduced from the three observed H$_2$CO lines (this study). Upper limits are given when the high-energy lines are not detected. The values tabulated when the three lines are detected should also be considered as upper limits, as the low-energy line is likely optically thick. \\
  $^d$ The lines have a non-Gaussian shape, the emission being dominated by the outflow. We derive therefore only an upper limit on the linewidth for the envelope, and refrain from giving a rotational temperature
  from H$_2$CO.\\
 $^e$ beam-averaged H$_2$ column densities in the 28$''$ APEX beam.\\
  $^f$ This value refers to the column density of the whole cloud, as traced by the C$^{18}$(3-2) line, corrected for opacity \cite{Bergman11a}. The central core, traced from H$_2$CO and CH$_3$OH lines, accounts for
  a column density of 3$\times$10$^{22}$\,cm$^{-2}$. This latter value was used to derive the 
  detected HOOH abundance \cite{Bergman11b}. \\
    \end{table}

{\bf G028.53-00.25 MM1} is the most massive core within the IRDC G028.53-00.25, another IRDC studied by \citet{Rathborne10}. 
They derive a temperature of 17\,K from their broadband SED study, a mass of 1088 M$_{\odot}$ from the 1.2\,mm data, and an H$_2$ column density of 5.6$\times$10$^{23}$ cm$^{-2}$ averaged in the MAMBO 11$''$ beam. Its angular size is 33$''$ \cite{Rathborne06}. This translates into an H$_2$ column density of 3.6$\times$10$^{23}$cm$^{-2}$ in a 28$''$ beam.

{\bf NGC6334I(N)} is a star formation site located north of the more developed NGC6334I ultracompact HII region and molecular core. \citet{Sandell00} mapped the region at five wavelengths in the range 350\,$\mu$m to 1.3\,mm. Although the temperature cannot be 
constrained independently from the dust opacity at these wavelengths, they found plausible fits in the temperature range $T$$_{\rm dust}$=30--35\,K. This is consistent with the temperature estimates of the extended gas by \citet{Kuiper95} (from low-energy NH$_3$ lines) and \citet{McCutcheon00} (from $^{12}$CO lines). Note that \citet{Beuther07} found evidence for higher gas temperature in the 
compact ($\sim$2$-$3$''$) core by means of NH$_3$ (5,5) and (6,6) inversion lines, but we are 
likely not sensitive to this hotter gas.
We derive $N$(H$_2$) = 4.2 $\times$ 10$^{24}$ cm$^{-2}$ averaged on the source from the average H$_2$ density and the source size (11$''$ $\times$ 8$''$) tabulated by \citet{Sandell00}. This translates into 
8.5$\times$10$^{23}$\,cm$^{-2}$ averaged in a 28$''$ beam.

{\bf G1.6$-$0.025} is one of the molecular clouds within the Central Molecular Zone (CMZ) surrounding the galactic center. It has been studied in detail by \citet{Menten09}. We targeted here the position 3 from 
\citet{Menten09}, for which they derived kinetic temperatures from a detailed analysis of CH$_3$OH excitation. The extended cloud, whose emission peaks at $v_{\rm lsr}$\,=\,51\,km\,s$^{-1}$, was shown to have a temperature of 60\,K, and an H$_2$ column density of 4$\times$10$^{22}$ cm$^{-2}$ (from $^{13}$CO measurements, in a 2$'$ beam) \cite{Menten09}. We assume here that this extended gas has the 
same column density at smaller scales. 

{\bf NGC1333 - IRAS4A} is a Class 0 low-mass protostar located in the NGC1333 
complex in the Perseus molecular cloud. The distance of this complex is 235\,pc \cite{Hirota08}.
At this distance, 28$''$ represent 6580 AU. We compute the H$_2$ column density averaged in this beam 
from the density power law derived by \citet{Kristensen12}, and find 1.3$\times$10$^{23}$ cm$^{-2}$.
\citet{Maret04} and \citet{Maret05} studied the emission of 
formaldehyde (H$_2$CO) and methanol (CH$_3$OH) towards this source, and derived rotational temperatures of 24\,K for both species.

{\bf L1527} is a young low-mass protostar located in the Taurus-Auriga molecular cloud, at a 
distance of 140 pc \cite{Kenyon94}. Its evolutionary stage is still debated \cite{Kenyon08}.
As for NGC1333-IRAS4A, we compute the H$_2$ column density from the density
power law of \citet{Kristensen12}. \citet{Maret04} derived a rotational temperature of 16\,K from the
study of H$_2$CO lines.

{\bf RCrA-IRS7B} is a Class 0 protostar located in the R Coronae Australis complex, at a distance
of 130 pc \cite{Neuhauser08}. Several molecular lines were observed using APEX towards this source by \citet{Schoier06}. The kinetic temperatures derived from H$_2$CO and CH$_3$OH lines are respectively 40 and 22\,K \cite{Schoier06}. Their study derives an H$_2$ column density of 3$\times$10$^{23}$ cm$^{-2}$ in the APEX2a 18$''$ beam, based on the C$^{34}$S line. \citet{Lindberg12} derived the density profile of the source, based on SCUBA and Herschel continuum data. Computation of the column density into a 18$''$ beam using their profile leads to 8$\times$10$^{22}$ cm$^{-2}$, which compared to the value derived from the C$^{34}$S line gives an idea of the uncertainty of the derivation of column densities from lines and dust (here a factor 4). Adopting the density profile\cite{Lindberg12}, we derive an H$_2$ column density of 5.9$\times$10$^{22}$ cm$^{-2}$ in a 28$''$ beam. 

\subsection{Observational results}

The HOOH $3_{0,3}- 2_{1,1}$ line was not detected towards any of the sources of our sample. The noise rms values reached towards each source are listed in Table \ref{tbl:results}. The rms levels reached are in the range of 16 -- 25\,mK  at 0.52\,km\,s$^{-1}$ resolution. 

Three H$_2$CO lines are present within the large bandwidth of the XFFTS. Their fluxes are also listed in Table \ref{tbl:results}, as they will be useful for the interpretation of the HOOH upper limits (Section \ref{Sect:analysis}). 

\begin{table}[!ht]
\small
\begin{centering}
  \caption{~Observational results}
  \label{tbl:results}
  \begin{tabular}{l l cccc} 
    \hline
    Source   &    HOOH     &    HOOH$^b$   &  H$_2$CO {\footnotesize 3$_{03}$-2$_{02}$} & H$_2$CO {\footnotesize 3$_{21}$-2$_{20}$} &  H$_2$CO {\footnotesize 3$_{22}$-2$_{21}$}   \\
    & & & {\footnotesize ($E_{\rm up}$\,=\,21.0\,K)} & {\footnotesize($E_{\rm up}$\,=\,68.2\,K)}&{\footnotesize($E_{\rm up}$\,=\,68.2\,K)} \\
         &     rms$^a$  &    $\int$ $T_{\rm a}dv$ & $\int$ $T_{\rm a}dv$ & $\int$ $T_{\rm a}dv$ & $\int$ $T_{\rm a}dv$ \\
    & mK  & mK\,km\,s$^{-1}$ & K\,km\,s$^{-1}$ & K\,km\,s$^{-1}$  & K\,km\,s$^{-1}$  \\
        \hline 
    \noalign{\smallskip}
 $\rho$ Oph-A$^c$  &  & 125 & 4.55$\pm$0.04 & 0.60$\pm$0.04 & 0.62$\pm$0.04 \\   
    \hline 
    \noalign{\smallskip}
$\rho$ Oph-B2-MM8           &     22.9 & $\le$\,45 &  1.83$\pm$0.02  &    $\le$0.02$^b$ & $\le$0.02$^b$ \\
    G15.01-0.67         &   19.2   &  $\le$\,92  & 20.57$\pm$0.03 & 5.64$\pm$0.03 & 5.28$\pm$0.03  \\ 
    G018.82-00.28MM1    &   21.1 & $\le$\,109 &  6.48$\pm$0.04 & 1.81$\pm$0.04 & 1.54$\pm$0.03 \\
    G018.82-00.28MM4    &   21.0 & $\le$\,96 &  0.87$\pm$0.03 & $\le$0.03$^b$ & $\le$0.03$^b$ \\
G028.53-00.25MM1   & 17.7   & $\le$\,96 &  2.17$\pm$0.03 & 0.61$\pm$0.03 & 0.42$\pm$0.04      \\
  
NGC6334I(N)         &   18.6 & $\le$\,97 &  44.18$\pm$0.06 & 18.94$\pm$0.04 & 17.72$\pm$0.07  \\
G1.6-0.025          &     16.1   & $\le$\,80 &  1.41$\pm$0.05 & 0.69$\pm$0.03 & 0.55$\pm$0.03 \\
NGC1333-IRAS4A       &   24.6& $\le$\,75 &  --$^d$ & --$^d$ & --$^d$   \\
L1527               &  18.4 & $\le$\,32 &  1.63$\pm$0.01 & 0.11$\pm$0.01 & 0.13$\pm$0.01   \\
RCrA-IRS7B           &  23.1    & $\le$\,61 &  9.21$\pm$0.02 & 1.94$\pm$0.02 & 1.84$\pm$0.02 \\
    \hline
  \end{tabular}\\
  \end{centering}
  $^a$ at resolution 0.52 km/s.\\
  $^b$ the tabulated upper limits on the integrated intensity are 3-sigma.\\
  $^c$ from \citet{Bergman11b} for HOOH and \citet{Bergman11a} for H$_2$CO.\\
  $^d$ see footnote $d$ of Table \ref{tbl:sourceproperties}
\end{table}

\section{Analysis}
\label{Sect:analysis}

\subsection{Upper limits on the HOOH column density}

In order to derive upper limits on the HOOH column densities from the rms noise of the observations, 
we need knowledge both of the typical linewidth in each source, as well as of the excitation
temperature of the considered HOOH transition.

The typical linewidth for each source can be inferred from the observations, as other lines are present
in the XFFTS range. For this purpose, we use the lower excitation line of H$_2$CO, which is detected towards
all sources, and derive its width using a Gaussian fit to the line. For sources where the higher energy 
lines of H$_2$CO are also detected, and the lower energy line is obviously broadened through 
opacity effects, we measure the width of the higher excitation lines. This linewidth is listed in the third column of Table \ref{tbl:sourceproperties}. The resulting 3$\sigma$ upper limits on the integrated flux are listed in column 3
of Table \ref{tbl:results}.

Estimating the excitation temperature of the line is more difficult. On the one hand, the dust temperature
has been measured for some of the sources (Table \ref{tbl:sourceproperties}). 
The dust temperature is expected to be equal to the gas temperature in dense regions where 
the two phases thermalize through collisions. But the average densities in our beam may not always be high 
enough to reach this state. 
 
On the other hand, we can compute the 
rotational temperature of H$_2$CO from the three para-lines which are in the observed band. 
In the case where the three lines are detected, this rotational temperature should be seen as an 
upper limit to the kinetic temperature, as the lower energy line is likely to be optically thick. 
We checked this approach on the SM1 core of Oph A (also called D-peak position) where HOOH was 
first detected \cite{Bergman11b}. We computed the $T_{\rm rot}$ based on only those three H$_2$CO lines, as observed by \citet{Bergman11a}. We find $T_{\rm rot}$\,=\,33$\pm$3 K, which is indeed higher than the kinetic temperature that was derived for this source from a more detailed modelling,
using more transitions and taking line opacities into account: 22.5\,K (modified rotation diagram technique) and 24\,K (ALI technique using H$_2$CO and CH$_3$OH)\cite{Bergman11a}. 

When the upper energy lines are not detected, we can likewise derive an upper limit on the rotational temperature from their non-detection. 

The excitation temperature of the HOOH line should lie somewhere in the range bracketed by the dust temperature and our derived H$_2$CO excitation temperature. 
In the cases where credible gas temperatures have been derived in previous studies (see column 5 of Table \ref{tbl:sourceproperties}), we take their values for the HOOH excitation temperature. This avoids using the likely overestimated 
H$_2$CO rotational temperature when the H$_2$CO lines are very optically thick.
This is for example the case of G15, where we get a 3$\sigma$ upper limit on HOOH abundance of 4$\times$10$^{-12}$ when assuming $T_{\rm ex}$\,=\,32\,K. Taking the likely overestimated temperature of 64\,K would lead to a value of 7$\times$10$^{-12}$. 
In cases where we do not have a good gas temperature estimate from the literature, we adopt the 
temperature we derived from H$_2$CO. This should result in an extremely conservative value (i.e. an 
overestimated upper limit), as described above. 

The 3$\sigma$ upper limits on the HOOH column densities are listed in Table \ref{tbl:analysis}. 
For the sake of clarity, the adopted $T_{\rm ex}$ for deriving the upper limit is also given. 
The 3$\sigma$ limits are of the same order or lower than the detected column density towards $\rho$\,Oph A, except for NGC6334(N) where the higher temperature pulled the limit on the column density towards a much higher value. This is consistent with the fact that the 1$\sigma$ upper limits of the integrated flux listed in Table \ref{tbl:results} are $3-10$ times lower than the detected flux towards $\rho$ Oph A.

\begin{table}[!h]
\small
\begin{centering}
  \caption{~(3$\sigma$) upper limits on the HOOH column densities and abundances relative to H$_2$}
  \label{tbl:analysis}
  \begin{tabular}{l ccc} 
    \hline
    Source   &    $T_{\rm ex}$   & ~~~~$N$(HOOH)~~~~   &   ~~ [HOOH]/[H$_2$] ~~   \\
        & K  &  cm$^{-2}$  &     \\
        \hline
        \noalign{\smallskip}
$\rho$ Oph-A & 22 & (3$-$8)\,$\times$10$^{12}$& (1$-$3)$\times$10$^{-10}$ $^{(a)}$\\
 &  & & (2$-$6)$\times$10$^{-11}$ $^{(b)}$\\
                \hline
        \noalign{\smallskip}
$\rho$ Oph-B2-MM8           &   16  & $\le$\,1.5$\times$10$^{12}$ & $\le$\,7$\times$10$^{-12}$\\
    G15.01-0.67         &   32 &  $\le$\,3.8$\times$10$^{12}$  & $\le$\,4$\times$10$^{-12}$ \\     
    G018.82-00.28MM1    &   61 &  $\le$\,8.1$\times$10$^{12}$ & $\le$\,5$\times$10$^{-11}$\\
    G018.82-00.28MM4    &  29 & $\le$\,3.7$\times$10$^{12}$  & $\le$\,4$\times$10$^{-11}$\\
    G028.53-00.25MM1A~~   & 57 &   $\le$\,6.7$\times$10$^{12}$ & $\le$\,2$\times$10$^{-11}$   \\
NGC6334I(N)         & 160 &  $\le$\,2.4$\times$10$^{13}$ & $\le$\,3$\times$10$^{-11}$  \\
G1.6-0.025            & 60 &$\le$\,5.9$\times$10$^{12}$&  $\le$\,1.5$\times$10$^{-10}$ \\
NGC1333-IRAS4A     &  24 & $\le$\,2.6$\times$10$^{12}$&  $\le$\,2$\times$10$^{-11}$ \\
L1527             & 16 & $\le$\,1.0$\times$10$^{12}$&  $\le$\,2.5$\times$10$^{-11}$  \\
RCrA-IRS7B         &  40 &  $\le$\,3.0$\times$10$^{12}$&  $\le$\,5$\times$10$^{-11}$\\
    \hline
            \noalign{\smallskip}
  \end{tabular}\\
  \end{centering}
The tabulated $T_{\rm ex}$ is the one used to derive the column densities. \\
$^{(a)}$ Using the compact core H$_2$ column density \cite{Bergman11a}. \\
$^{(b)}$ Using the full H$_2$ column density. 
\end{table}

\subsection{HOOH abundances}

We derived the upper limits for the HOOH abundance in each source, by using the H$_2$ column density
averaged over the APEX beam (see section \ref{sources} for detailed discussion for each source). The resulting 3$\sigma$ upper limits are tabulated in Table~\ref{tbl:analysis}. We revisited the case of Oph A, by estimating the abundance of HOOH under the assumption that  the emission comes from the full gas along the line-of-sight (case\,(b) in Table \ref{tbl:analysis}, where the full H$_2$ column density is traced from C$^{18}$O(3-2) observations), while \citet{Bergman11b} assumed it originated in the dense core traced in H$_2$CO and CH$_3$OH (case (a) in Table \ref{tbl:analysis}).

All 3$\sigma$ upper limits (except for the case of G1.6, where the relatively high upper limit stems from a combination of high excitation temperature and low H$_2$ column density) are well under the detected abundance of HOOH in OphA, when the detected molecule is assumed to be located in the compact core \cite{Bergman11b}. All 3$\sigma$ upper limits are of the same order as or slightly lower than the HOOH abundance in Oph A, if the full column density traced by C$^{18}$O(3-2) is taken into account. 

The derived upper limits are therefore significant, and constrain the abundance of HOOH to a strictly lower value than in $\rho$ Oph A in all sources.

\section{Discussion}

The main result of our observational search is that HOOH is very rare in the interstellar medium. This sheds some new light on why the molecule was only detected very recently\cite{Bergman11b}.
Under the specific physical conditions of $\rho$ Oph A, HOOH is however abundant, so the puzzle remains to understand what makes this source so different from all other sources from our sample. 

In the following, we discuss the direct implications of our observations, as well as the implications in terms of
chemical modelling.

\subsection{HOOH and O$_2$, similarly elusive}

We note that the detection of HOOH towards one sole source ($\rho$ Oph A) is very similar to 
the case of O$_2$, which has been searched towards many sources \cite{Goldsmith00, Pagani03}, and detected so far in only two sources\cite{Larsson07,Liseau12,Goldsmith11}, the strongest case being $\rho$ Oph A, for which the O$_2$ abundance relative to H$_2$ is\cite{Liseau12} 5$\times$10$^{-8}$. It is certainly not a simple coincidence that O$_2$ and HOOH have been detected towards the same source, and that they are both otherwise elusive, as their chemistry is tightly linked via the grain surface reactions (1) and (2). 

\subsection{The role of the environment} 

Our source sample allows us to investigate the role of the environment and of possible local chemical anomalies 
on the abundance of HOOH. $\rho$ Oph B belongs to the same molecular cloud as $\rho$ Oph A, and is situated
therefore also at the same short distance from the sun (120\,pc). 
The H$_2$ column density that we derive for $\rho$ Oph B is somewhat higher than that of $\rho$ Oph A. 
The 3$\sigma$ upper limit on the HOOH abundance in $\rho$ Oph B is very low (3 to 50 times lower, depending on 
the assumption for the location of HOOH in $\rho$ Oph A) compared to the HOOH abundance observed in 
$\rho$ Oph A. 

The lack of detection towards $\rho$ Oph B seems to discard the possibility that the detection in $\rho$ Oph A is 
due to anomalous initial elemental abundances.   
Instead, the comparison of these two sources seems to indicate that the temperature is important: $\rho$ Oph B is
rather cold ($\le$16\,K), while $\rho$ Oph A has a temperature of 24\,K. We address this point in more detail in the following sections. 

\subsection{The role of the present average temperature and density conditions}

\citet{Du12} modelled in detail the abundance of HOOH in Oph A, as well as that of many other observed molecules also believed to form on dust grains, using a fully coupled gas-grain model, taking into account the latest experimental results \cite{Cuppen10} for the reaction rates on the grains. Their model was able to reproduce the abundance of HOOH in the source, and even predicted successfully the abundance of O$_2$H, which was detected subsequently \cite{Parise12a}. Gas-phase HOOH was found to originate mainly from the desorption of HOOH formed on the grains through reaction (2). 
The model of \citet{Du12} assumed a constant temperature of 21\,K, and a constant density of 6$\times$10$^5$\,cm$^{-3}$. The best match for the abundance of all considered observed molecules (O$_2$, HOOH, HO$_2$, H$_2$CO and CH$_3$OH) was obtained for an age of 6$\times$10$^5$ yrs. At earlier times, HOOH was found to be overabundant. 
They showed that the temperature plays an important role in the HOOH abundance, whereas the role of the density is not as significant. 
A change from 20 to 22\,K was shown to increase the HOOH abundance by an order of magnitude, while the abundance did not vary much between 22\,K and 30\,K (see their Figure 4). At the given evolutionary time of 6$\times$10$^5$ yrs, the abundance of HOOH decreased with increasing density in the range $10^5-10^6$ cm$^{-3}$ (see their Figure 6). Their modelling implies that {\it the HOOH abundance should be at least as high as in Oph\,A for sources younger than 6$\times$10$^5$ yrs, with temperatures in the range 21$-$30\,K and densities lower than or equal to 6$\times$10$^5$ cm$^{-3}$} (in the assumption of constant temperature and density).

In our sample, several sources have an average temperature in the range 21$-$30\,K. Among them, IRAS4A is certainly the best source to compare to Oph\,A (although IRAS4A is almost twice as distant, cf. Table \ref{tbl:sourcelist}). Its present temperature is similar to that of Oph A. An LVG study of the H$_2$CO emission \cite{Maret04} led to an estimated density of (3$-$4)$\times$10$^{5}$ cm$^{-3}$, while the 
density at a radius equal FPBW/2 is $8\times10^5$ cm$^{-3}$
according to the density profile of \citet{Kristensen12}. The average density of IRAS4A is therefore also very similar to that of Oph A. 
Unless this former source is older than (6$-$10)$\times$10$^5$ yrs, the model of \citet{Du12} would therefore predict an HOOH abundance at least as high in this source as in Oph A, within the assumption of stationary temperature and density. The lack of detection of HOOH towards this source is therefore puzzling.\\

The low-mass Class 0 protostar IRAS4A was much colder in the past, having, as already pointed out by \citet{Yildiz13}, likely evolved through a long ($\ge$ 8$\times$10$^5$ yrs) cold precollapse phase (T\,$\sim$\,10\,K). 
\citet{Yildiz13} derived a low 3$\sigma$ upper limit for the abundance of O$_2$ towards NGC1333- IRAS4A, based on Herschel observations. Their interpretation for the low abundance of O$_2$ is that most of the O$_2$ formed in the gas at early times is hydrogenated into water on the grains during the long cold precollapse phase. As the hydrogenation of O$_2$ leads to the formation of HOOH, our non-detection of HOOH might bring some new constraints on this interpretation. Furthermore, their study shows that the temporal evolution of the physical
conditions is playing an important role in the non-detection of O$_2$.

In order to investigate further what could be the key difference between Oph A and the other sources, 
we performed new chemical model calculations, including models with non-stationary temperature and density. 

\subsection{Chemical modelling}

{\bf a. Description of the chemical model} 
\vspace{0.2cm}\\
The model is based on the same assumptions as that of \citet{Du12}, and takes into account the correction from \citet{Du12b}. The model has been described in detail in \citet{Du12}, and we summarize here its main characteristics. It solves the coupled gas-phase and grain surface chemistry, using the Hybrid Moment Equation (HME) method \cite{Du11}, which was developed to address correctly the stochasticity of the grain surface chemistry. The HME code was benchmarked against Monte Carlo simulations, and showed very good agreement in the results\cite{Du11}. No layering of the ices is considered in this version of the code. The surface reactions considered in the model are listed in appendix B of \citet{Du12} (but see also \citet{Du12b}). They are based on a combination of selected reactions from \citet{Allen77}, \citet{Tielens82} and \citet{Hasegawa92}. Some of these earlier reaction rates have been updated according to recent experimental results \cite{Ioppolo08,Fuchs09,Cuppen10,Ioppolo10}. We assume quantum tunneling for the reactions having a barrier. The reaction probabilities depend on the product $ a \sqrt{E_a} $, where $a$ is the barrier width, 
and $E_a$ the barrier height. The absolute values of these parameters are therefore not needed, and we can without
introducing any limitation for our purposes assume $a$\,=\,1\,\AA~for all reactions, and then derive the corresponding barrier height $E_a$ from the experimental results. In particular, the barrier heights of reactions (2) and (3) are estimated based on \citet{Cuppen10}, while that of reaction (1) 
is assumed to be 600\,K (a value which is intermediate between those of \citet{Tielens82} and \citet{Ioppolo10}). 
Photodissociation reactions induced by cosmic rays and chemical desorption reactions are also included. 
\vspace{0.2cm}\\
{\bf b. Stationary physical conditions} 
\vspace{0.2cm}\\
The time evolution of physical parameters in observed sources is in general very difficult to constrain. For example, it is unknown how steep the increase of temperature is during the formation of a protostar. It is therefore useful in a first step to look at chemical models with stationary physical conditions, to understand the first order effects of the temperature and density on the chemistry.

\begin{figure}[!ht]
\begin{center}
\begin{tabular}{cl}
\includegraphics[width=0.48\textwidth, trim= 0.5cm 2cm 2.0cm 3.5cm , clip=true]{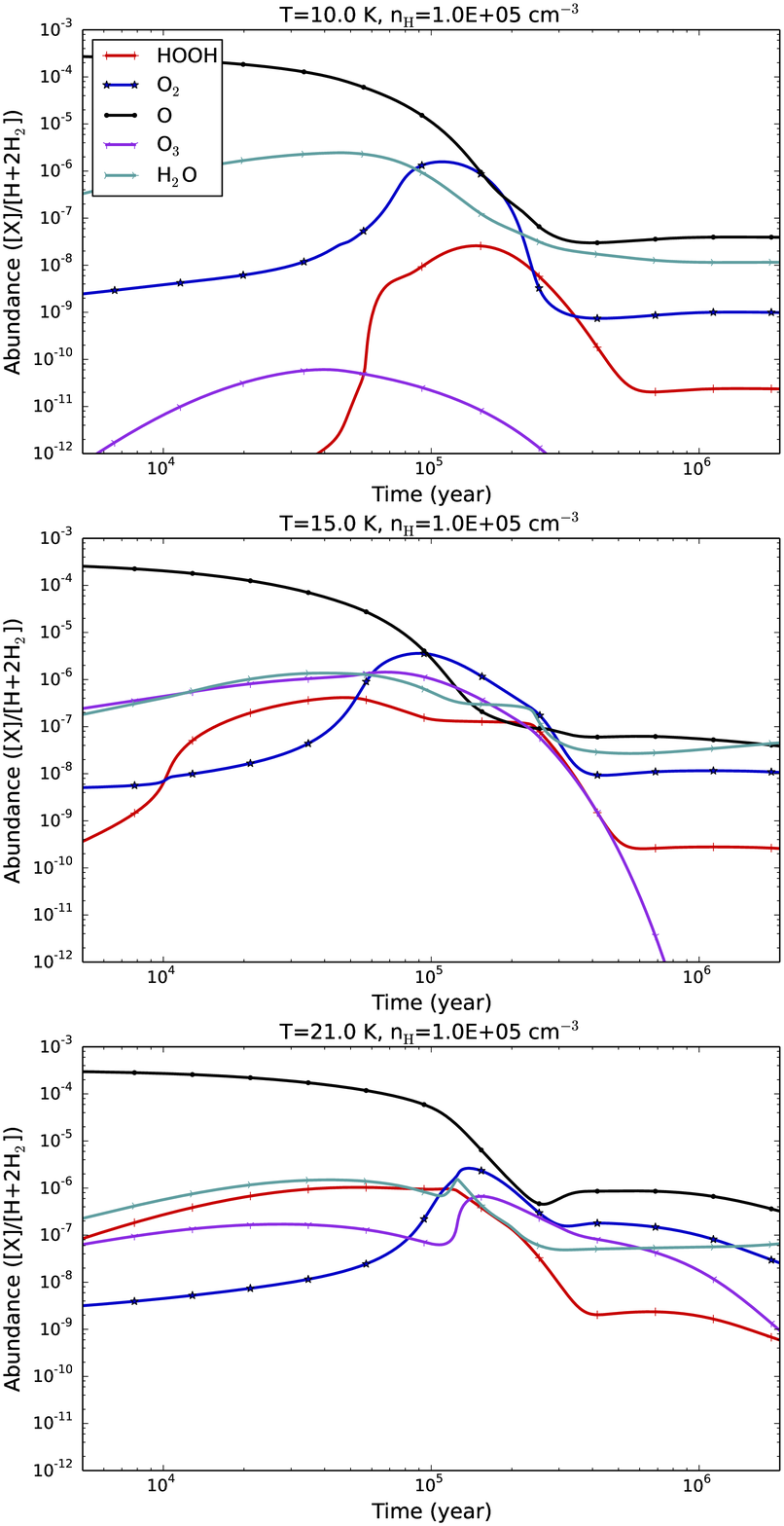} & \includegraphics[width=0.48\textwidth, trim= 0.5cm 2cm 2.0cm 3.5cm, clip=true]{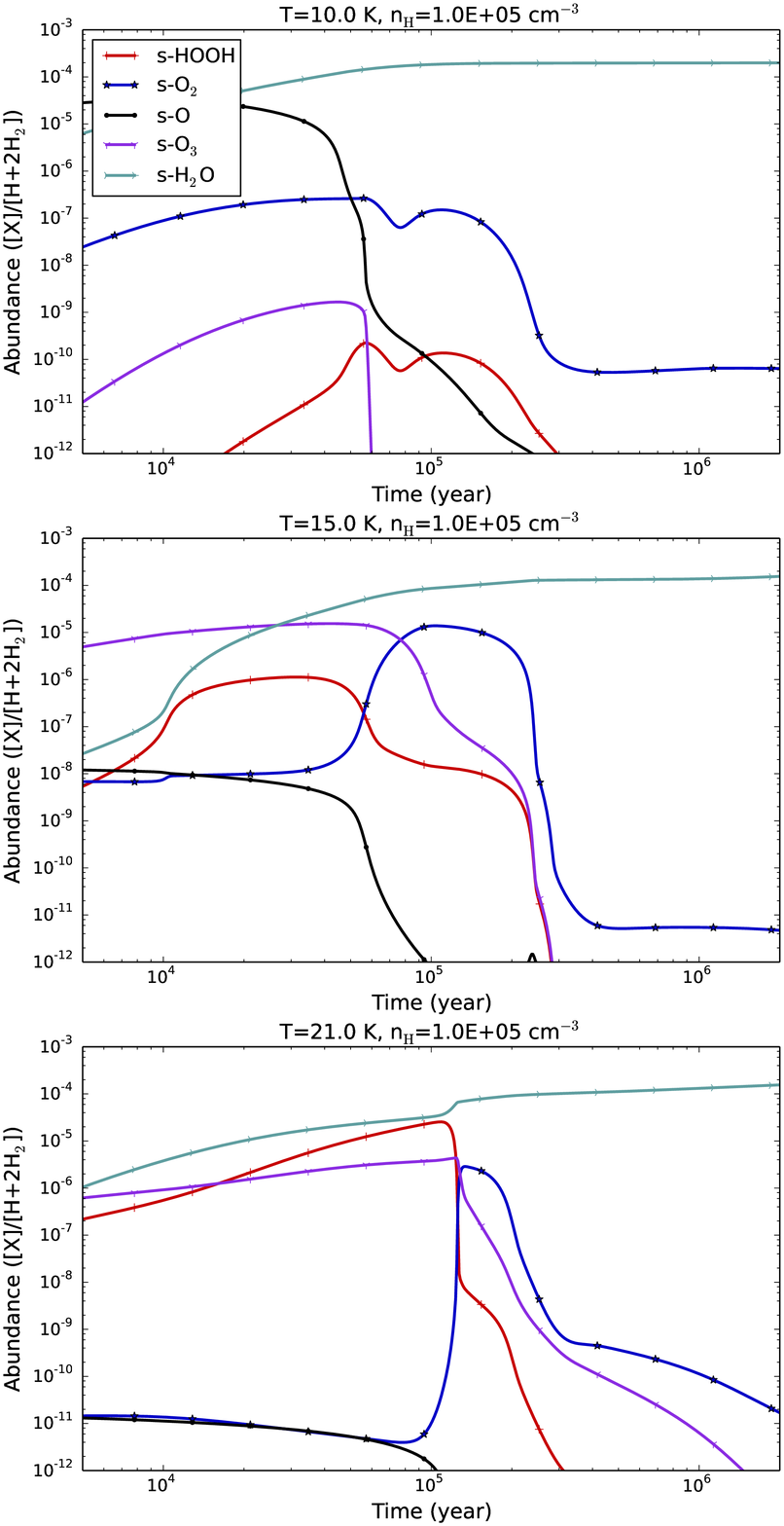}\\
\end{tabular}
\caption{Evolution with time of the gas-phase (left panel) and solid (ice, right panel) abundances of O, O$_2$, O$_3$, HOOH and H$_2$O for the fixed density of 10$^5$ cm$^{-3}$, and temperatures of 10\,K (upper panels), 15\,K (central panels) and 21\,K (lower panels).}
\label{HME}
\end{center}
\end{figure}

Figure \ref{HME} shows the predictions for three different temperatures (10\,K, 15\,K and 21\,K), other parameters being identical. We fixed the density to 10$^5$ cm$^{-3}$ (the value chosen by \citet{Yildiz13} for the precollapse
phase for IRAS4A). The predicted abundance of HOOH at late times is increasing with temperature in the presented range, as is the abundance of O$_2$ (as already noticed by \citet{Du12} at a higher density). 
This might already give the reason for the abundance of HOOH in Oph B (at temperature 12\,K) being lower than 
in Oph A. Our model predicts that HOOH is more than an order of magnitude less abundant at 12\,K than at 21\,K at 
high times.

It is interesting to notice that O$_2$ forms on grain surfaces at 10\,K in our model, even within our assumption that the diffusion barrier is a relatively high fraction (0.77) of the desorption barrier compared to other models (\citet{Yildiz13} used 0.5). This surface formation outweights the contribution of the freeze-out of gas-phase O$_2$ to the abundance of O$_2$-ice until about 10$^5$ years.  This is in contradiction with the conclusion one can reach based on the very low mobility of oxygen atoms compared to H atoms at these temperatures. This might be due to a relatively high accreting O/H ratio at the beginning of the evolution, leading to a high concentration of O on the surface. 

At early times, the HOOH abundance is higher than at t\,$> 6\times10^5$yrs. 
As \citet{Du12} already mentioned, the gaseous destruction mechanisms for HOOH are likely not complete in the present chemical networks, and therefore the HOOH abundance may be overall overestimated.
It is therefore possible that the rarity of HOOH is due to an observational bias, and that Oph A corresponds to the rare example of a young object. Such an interpretation could only be confirmed after the gas-phase destruction mechanisms of HOOH 
have been reviewed, and the thorough modelling study from \citet{Du12} is updated, to take into account again the 
agreement of the abundance of many species simultaneously. From now on, we will focus on other possible interpretations for the detection of HOOH in Oph\,A.
\vspace{0.2cm}\\
{\bf c. IRAS4A: is HOOH consistent with the O$_2$ upper limit?}
\vspace{0.2cm}\\
We can now turn back to the problem of HOOH in IRAS4A. Our model with $T$\,=\,10\,K and $n_{\rm H}$\,=\,10$^5$ cm$^{-3}$, corresponds to the conditions assumed by \citet{Yildiz13} for the precollapse phase of IRAS4A.

Our model predicts an [O$_2$]/[H+2H$_2$] abundance ratio of $\sim$ 10$^{-9}$ (i.e. 
[O$_2$]/[H$_2$] $\sim$2$\times 10^{-9}$ at times longer than 8$\times$10$^{5}$ yrs (time beyond which the
model settles to constant values), which is consistent 
with the upper limit on the O$_2$ abundance towards IRAS4A\cite{Yildiz13}. 

The predicted abundance of HOOH is [HOOH]/[H$_2$]\,$\sim$\,5$\times$10$^{-11}$. This is 
barely consistent with our non-detection ([HOOH]/[H$_2$]\,$\le$ 2$\times$10$^{-11}$, 3$\sigma$). 
A slightly higher density has the effect to lower the expected HOOH abundance (for $n_{\rm H}$\,=\,2$\times$10$^5$ cm$^{-3}$, the predicted abundance decreases to 3$\times$10$^{-11}$). 
Here again, the model might be reconciled with the observations if missing gaseous destruction mechanisms for 
HOOH were added. A more detailed test would involve doing a shell modelling as \citet{Yildiz13} did, to account for the 
increased density and temperature after the embedded protostar formed, but this is beyond the scope of this paper.
\vspace{0.2cm}\\
{\bf d. The warming-up phase}
\vspace{0.2cm}\\
We want to further investigate the role of the temperature in the formation of HOOH. However, one cannot realistically model the conditions of star-forming regions using stationary conditions. Indeed, star-forming regions are 
evolving from cold cloud conditions to warm conditions. At the start, cold conditions ensure that grain-surface chemistry can play a key role, whereas it would not in models which already start with warm conditions. We here ran chemical models with a warming-up phase, with the aim to further inquire the impact of the temperature on the abundance of HOOH.  

\begin{figure}[!ht]
\begin{center}
\includegraphics[width=0.99\textwidth, trim= 0.5cm 1cm 2.0cm 2cm , clip=true]{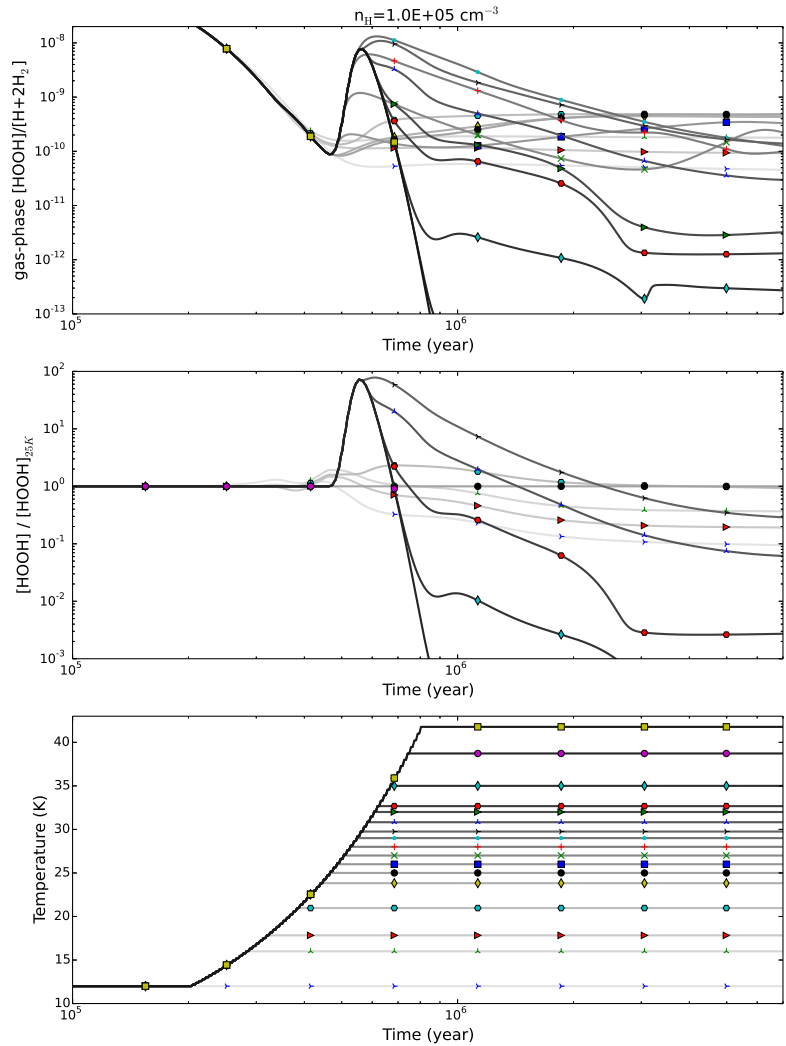}
\caption{Evolution with time of the abundance of gas-phase HOOH (upper panel), for the warming-up models. The temperature evolution is shown in the lower panel, while the middel panel shows the abundance of HOOH normalized to that in the 25\,K model. }
\label{HME2}
\end{center}
\end{figure}

All models have a constant density of 10$^5$ cm$^{-3}$. The early-time temperature is taken to be 12\,K, and the warm-up phase starts at 2$\times$10$^5$ yrs. The temperature then increases by 1\,K every 2$\times$10$^4$ yrs. The models differ from each other by the temperature at which the warm-up phase stops. Then the temperature is being kept constant until the end of the evolution (10$^7$ yrs). The resulting HOOH abundances are shown in Figure \ref{HME2}.

The abundance of HOOH for the model that remains at 12\,K has a similar behaviour as that of the 10\,K model (from 
the previous section), but with a final steady-state abundance about twice as high as at 10\,K. When increasing the temperature
from 12\,K, the HOOH abundance increases in the first place, and stays at all times higher than that of the 12\,K model, 
as long as the reached temperature remains under 30\,K. In the case of a final temperature of 21K, the HOOH abundance is more than one order of magnitude larger than that at 12\,K at all times. For final temperatures of 28 to 30\,K, the enhancement of the HOOH abundance can reach up to two orders of magnitude, and slowly decays again to less than one 
order of magnitude enhancement in a few 10$^6$ yrs.

For sources in which the warm-up reaches temperatures above 31\,K, the situation changes though. The decrease of the HOOH abundance is steep very soon after the initial enhancement appearing around 28$-$30\,K. For T\,$>$\,33\,K, the abundance decreases fast to values several orders of magnitude lower than the abundance for 12\,K.  

We also ran similar models, for which the warm-up phase started later (at 7$\times$10$^5$ yrs), and found that the main
facts listed above still apply, and that the H$_2$O$_2$ abundance reached at later times is unchanged. The duration of the cold phase therefore does not seem to be of much significance.

The conclusion of this study is that enhancement of the HOOH abundance does occur is a very limited range of temperatures, around 20$-$30\,K. Any further warm-up above 30\,K will result in the rapid destruction of HOOH.

\subsection{The special case of $\rho$ Oph A}

The detection of HOOH in Oph A in stark contrast to the non-detections in the other sources could therefore
reflect that the bulk of the material in Oph A is within the favorable temperature range (20$-$30\,K), whereas
this is not the case for the other sources. In Oph A, the estimates of the gas temperature at different positions
of the clump point to temperatures in the range 24$-$30\,K (see e.g. Tab. 8 from \citet{Bergman11a}). 
The fraction of the mass in the 20$-$30\,K range to the total mass may be close to unity.

For the IRAS4A protostar, the average temperature is 24\,K, as for Oph A. But here the protostar is heating internally
its envelope, causing a steep temperature gradient. As a result, only a small portion of the gas is actually at the 
average temperature. Using the density profile from \citet{Kristensen12}, and approximating the temperature profile
with a power law, we find that only $\sim$\,15\% of the total envelope mass in IRAS4A is in the temperature range 20$-$30\,K. Therefore, the HOOH abundance enhancement should be strongly suppressed in this object. Considering only the 
H$_2$ in the range 20$-$30K, the upper limit on the HOOH abundance in this gas becomes 1.3 $\times$ 10$^{-10}$\,cm$^{-3}$, 
a value that does not conflict anymore with the detection in Oph A. This interpretation could be further put 
to test by integrating much deeper on IRAS4A.

The other sources in our sample may similarly not have the bulk of their mass in the 20$-$30\,K range. 
The IRDC sources
might be warmer than $\sim$\,35\,K (as suggested by the high rotational temperatures for H$_2$CO), in which case the predicted abundance for HOOH would fall to a few 10$^{-12}$ 
or even lower. Additionally, these sources are much further away, so that the material that might
be at the favorable temperature is heavily beam-diluted.
On the contrary, Oph B is too cold to have a significant enhancement of HOOH. Our model with stationary conditions $T$\,=\,12\,K and $n$\,=10$^6$ cm$^{-3}$ predicts an HOOH abundance of $\sim$ 2 $\times$ 10$^{-12}$ at late times, a value well
below our observed upper limit. 
Finally, the case of the two other low-mass protostars L1527 and RCrA$-$IRS7B is certainly similar to that of IRAS4A, where the internal heating by the protostar may cause a steep temperature gradient. 

How could $\rho$ Oph A achieve this particular condition? A closer look at the environment of Oph A 
shows that it is externally heated by the S1 source\cite{Liseau99}, which is in fact a close binary system (B4 + K)\cite{Gagne04}. The slightly curved morphology to the East of the main ridge of Oph A (see e.g. Figure 1 
from \citet{Larsson07}) seems to coincide with the edge of the ISOCAM bright emission surrounding the S1 source
\cite{Abergel96}. This might be the sign that Oph A was formed from compression under the radiative pressure from the B4 star. Compression and external heating may be the cause of the unusual warm conditions within Oph A.

We finally remark that, because of the tight chemical link between O$_2$ and HOOH, we expect the abundance of both molecules to show some degree of correlation in astronomical sources. At longer times ($t>8 \times10^5$ yrs), we get from our (static) chemical models [O$_2$]/[HOOH] $\sim$ 40 for a range of temperatures between 10 to 21\,K, at the fixed density of 10$^5$\,cm$^{-3}$. The search for O$_2$ has proven to be very difficult, because of the atmospheric opacity at the frequencies of its intrinsically weak magnetic dipole rotational transitions, 
requiring deep observations from satellites (SWAS, Odin, Herschel). We suggest here that the search for sources with high O$_2$ content could be easily approached by searching first for HOOH, a molecule much easier to target with ground-based telescopes.

\section{Conclusions}

Following up on our detection of HOOH towards Oph A, we have searched for HOOH in a sample of ten sources, of different nature and in different environments. HOOH was not detected towards any of the sources in our sample, and significant upper limits for the HOOH abundance could be obtained. These negative results shed new light on the key parameters in the O$_2$/HOOH chemistry. We ran new gas-grain chemical models, taking into account a warm-up phase after a cold early cloud phase. The models show that the production of HOOH is extremely sensitive to the temperature, and that outside 
of the 20$-$30\,K temperature range the expected HOOH abundance is very low. We conclude that the key difference between Oph A and the other sources is that the bulk of the material in Oph A is likely to be at this favorable temperature,
whereas most of the mass may lie outside this range for the other sources. This peculiar condition for Oph A may be caused by external heating. This interpretation could explain the scarcity of detections of O$_2$ and HOOH in the ISM, and could 
be tested by observing other externally heated sources. 

\section*{Acknowledgements}
We thank F. Du for interesting discussions during the early phase of this project, as well as R. Liseau for discussions regarding the stellar environment of $\rho$ Oph~A.
\footnotesize{
\bibliography{/Users/bparise_back/These/Manuscrit/biblio} %your .bib file
%\bibliography{rsc}
\bibliographystyle{rsc}
}

\end{document}